# A unified model of particle mass

**Bernard F. Riley**

NNC Ltd., Birchwood Park, Warrington WA3 6BZ, United Kingdom.
bernard.riley@nnc.co.uk

**Abstract**

The quark masses evaluated by the Particle Data Group are consistent with terms in a geometric progression of mass values descending from the Planck Mass. The common ratio of the sequence is $2/\pi$. The quarks occupy the 'principal' levels of the mass spectrum, characterised by a new quantum number, n. Hadrons occupy mass sub-levels, characterised by fractional values of n. The quark masses of the model are used to formulate hadron mass construction equations based on the masses of neutral precursor particles. Hadron mass partnerships occur, in which mass differences arising from differences in spin, isospin and isospin projection are equal to the masses of principal levels. Mass partnerships also occur between the charged leptons and pseudoscalar mesons.

## 1 Introduction

In the Standard Model of particle physics [1], the Higgs mechanism of spontaneous electro-weak symmetry breaking gives the particles their masses. Precision electro-weak measurements [2] suggest that the mass of the Higgs boson is less than a few hundred GeV. But because quantum corrections to the Higgs mass are quadratically divergent in the Standard Model, the Higgs boson can receive a very large radiative mass correction from a higher mass scale, which is inconsistent with the requirement that its mass should be relatively low. This is the 'hierarchy problem' of particle physics, equivalent to explaining the immense gulf in mass between the electro-weak scale (~$10^3$ GeV) and the GUT scale of ~$10^{16}$ GeV or the Planck scale of ~$10^{19}$ GeV.

Various approaches have been developed to resolve the hierarchy problem. Among these are supersymmetry [3], technicolour theories [4] and extra-dimensional theories [5-10]. Whatever the solution turns out to be, it should eventually be incorporated into an all-encompassing theory that includes gravity and reconciles it with quantum mechanics [11]. The unification mass of the all-encompassing theory would be of the order of the Planck scale, the natural scale of string theory. According to string theory, the properties of an elementary particle, including its mass, are determined by the resonant pattern of vibration that its internal string executes [12]. The energy of the vibrating string is largely cancelled by quantum mechanical effects to produce the comparatively minute masses of the particles identified to date.

The unified model has been built upon the premise that particle masses are related directly to the Planck Mass. The quarks of the Standard Model are conjectured to be the elementary units of the model, occupying levels within a mass spectrum that descends from the Planck Mass in a geometric progression. The mass of the nth level is given by

$$m_n = k^{-n} m_p \qquad (1)$$

where k is a constant and $m_P = (\hbar c / G)^{1/2}$ is the Planck Mass.

First, the value of k in (1) is determined, from which values of n are deduced for the quarks. Non-integer values of n are then presented for the hadrons of zero orbital angular momentum in the $J^P={}^1/_2{}^+$ light baryon octet, the $J^P={}^3/_2{}^+$ light baryon decuplet, the pseudoscalar and vector light meson nonets, for charmed and bottom hadrons of zero orbital angular momentum and for charmonium and bottomium states. Building on the values of quark mass generated by the analysis, hadron mass construction equations are formulated. Finally, particle mass partnerships are identified. All mass measurements referred to in this paper have been taken from the Particle Data Group's listings, 2002 [13].

## 2 The Particle Mass Equation

To determine the value of k in (1), relationships were sought between the quark masses. The Particle Data Group evaluate a range of mass values for each quark, so it was necessary to extract from the ranges a single value of mass for each quark. The mid-range values were taken. The data are presented in Table 1.

**Table 1 Quark mass evaluations of the Particle Data Group, 2002 [13]**

| Quark | Mass evaluation range | Mid-range value |
|---|---|---|
| up (u) | 1.5 - 4.5 MeV | 3 MeV |
| down (d) | 5.0 - 8.5 MeV | 6.75 MeV |
| strange (s) | 80 - 155 MeV | 117.5 MeV |
| charm (c) | 1.0 - 1.4 GeV | 1.2 GeV |
| bottom (b) | 4.0 - 4.5 GeV | 4.25 GeV |
| top (t) | 174.3 ± 5.1 GeV | 174.3 GeV |

The up, down and strange quark masses of Table 1 are $\overline{MS}$ values at a scale $\mu \approx 2$ GeV. The charm and bottom quark masses are 'running' masses in the $\overline{MS}$ scheme. The top quark mass results from direct observations of top events.

Since a wide range of values is found for the mass ratios of those quarks adjacent in mass, the quarks will not occupy consecutive levels in a mass spectrum of the form given in (1). The $m_b : m_c$ ratio of approximately 3.5 ± 1 : 1 suggests that k in (1) is roughly equal to 3.5 or is a root of that number, perhaps close to 1.9 or 1.5. Values of k were tested to identify a value that would produce near-integer values of n for the mid-range values of quark mass evaluation. Crucial, though, in the identification of k was the search for patterns among the values of n calculated for hadrons. Such patterns did not begin to emerge among any group of hadrons, e.g. baryon singlet states, until k was set to a value within 1 part in $10^4$ of $\pi/2$.

In Figure 1, the values of n corresponding to the quark mass evaluation ranges of the Particle Data Group are shown superimposed upon the mass level spectrum resulting from the particle mass equation

$$m_n = \left(\frac{\pi}{2}\right)^{-n} m_P \qquad (2)$$

The mid-range quark mass values lie close to the mass levels resulting from (2). The indicated values of integer n are inserted into (2) to produce the quark masses of the model. The values are presented in Table 2.

**Table 2 Quark masses corresponding to levels in the mass spectrum**

| Quark | Mass level number, n | Mass | Mid-range value of Particle Data Group evaluation |
|---|---|---|---|
| up | 110 | 3.262 MeV | 3 MeV |
| down | 108 | 8.050 MeV | 6.75 MeV |
| strange | 102 | 120.9 MeV | 117.5 MeV |
| charm | 97 | 1.156 GeV | 1.2 GeV |
| bottom | 94 | 4.482 GeV | 4.25 GeV |
| top | 86 | 166.1 GeV | 174.3 GeV |

The mass of 166.1 GeV for the top quark is close to the usual top quark $\overline{MS}$ mass, $\overline{m}_t(\overline{m}_t)$ = 165 GeV [14].

The quark mass level numbers for the three families of the Standard Model are presented in Figure 2, which highlights the relationships between the least massive quarks of the three families of particles and the relationships between the quarks of each family. These relationships are described by the equations:

$$\frac{m_s}{m_u} = \frac{m_b}{m_s} = \left(\frac{\pi}{2}\right)^8 \qquad (3)$$

$$\frac{m_d}{m_u} = \left(\frac{\pi}{2}\right)^2 ; \frac{m_c}{m_s} = \left(\frac{\pi}{2}\right)^5 ; \frac{m_t}{m_b} = \left(\frac{\pi}{2}\right)^8 \qquad (4)$$

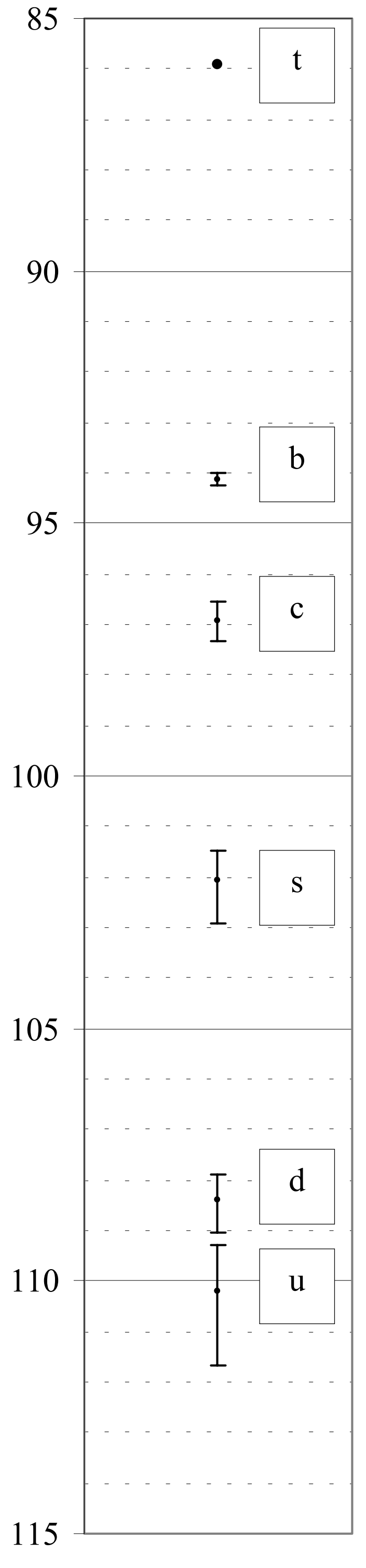

Figure 1: Mass level numbers, n, of the quark mass evaluations of the Particle Data Group [13] superimposed upon the principal levels of the mass spectrum.

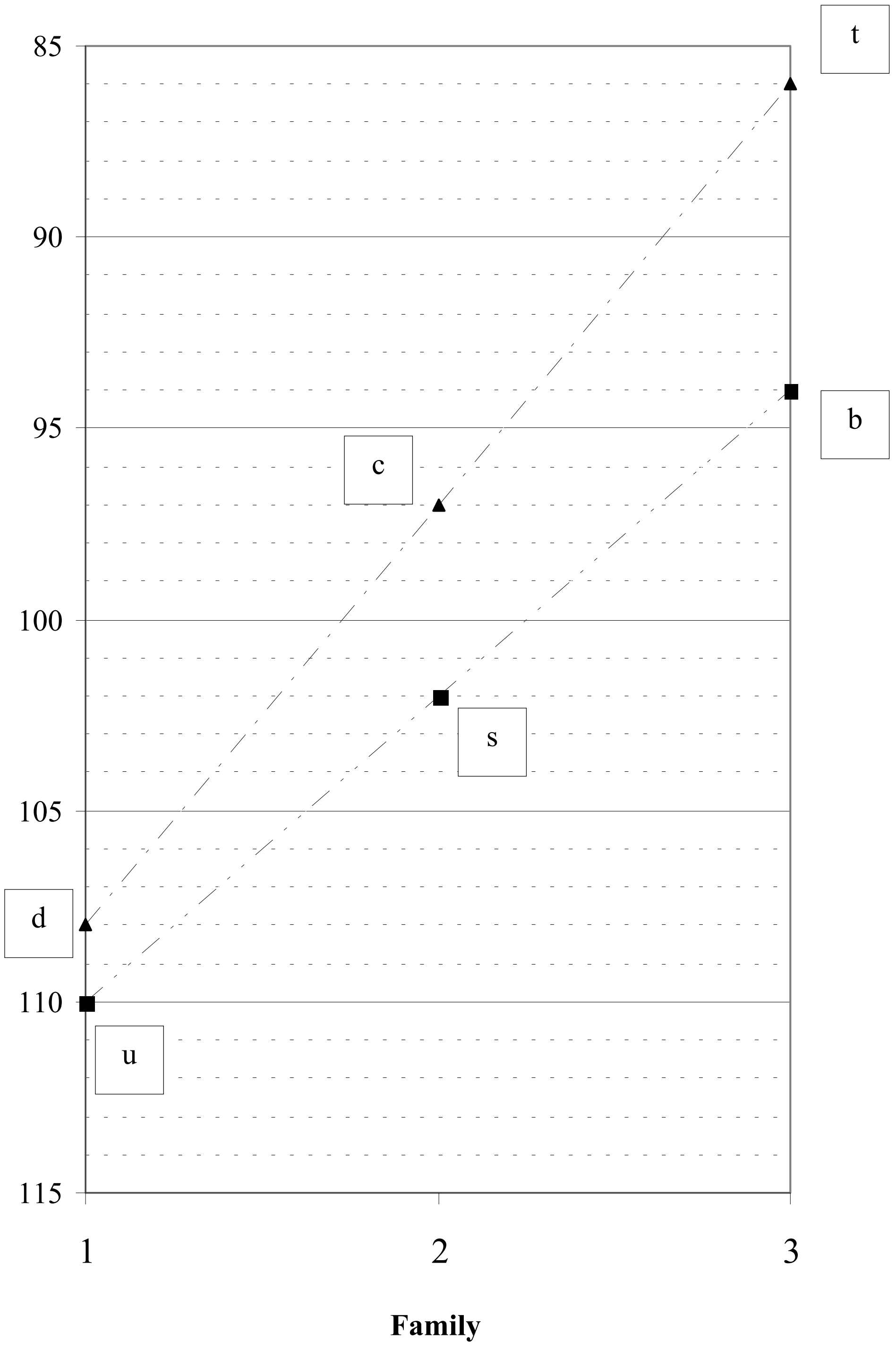

Figure 2: Mass level numbers, n, of the quarks of the Standard Model.

Integer values of n in (2) are characteristic of 'principal' mass levels. Hadrons occupy mass sub-levels characterised by non-integer values of n. The masses of the first order sub-levels are given by n = r/2 in (2), where r is an integer. The masses of yth order sub-levels are given by $n = r/2^y$.

Uncertainty in the value of the Planck Mass ($1.2210 \pm 0.0009 \times 10^{19}$ GeV) gives rise to a systematic relative uncertainty of $7 \times 10^{-4}$ in the masses of all levels. For mass sub-levels of fifth order, this uncertainty is equal to 0.024 of the level spacing and is usually greater than the particle measurement uncertainty. Planck Mass uncertainty and particle mass measurement uncertainty will therefore not obscure particle mass patterns involving low order mass sub-levels. The effect of uncertainty in the Planck Mass will be considered further when considering the high order mass sub-levels occupied by mesons. Throughout the paper, measurement uncertainties are displayed on graphs as error bars wherever their sizes are significantly greater than the diameters of the data point markers.

## 3 Hadron mass levels

The values of n for all light unflavoured and strange hadrons from the Particle Data Group listings lie between 95 and 102. The n-values of all charmed and bottom hadrons, and $c\bar{c}$ and $b\bar{b}$ states, have been superimposed upon the mass level spectrum in Figure 3. All particles in each of these four categories are confined to an integer level interval. The n-values of the charmed hadrons lie between 95 and 96 while those of the $c\bar{c}$ states lie between 94 and 95. Values of n for the bottom hadrons lie between 93 and 94 while those of the $b\bar{b}$ states lie between 92 and 93.

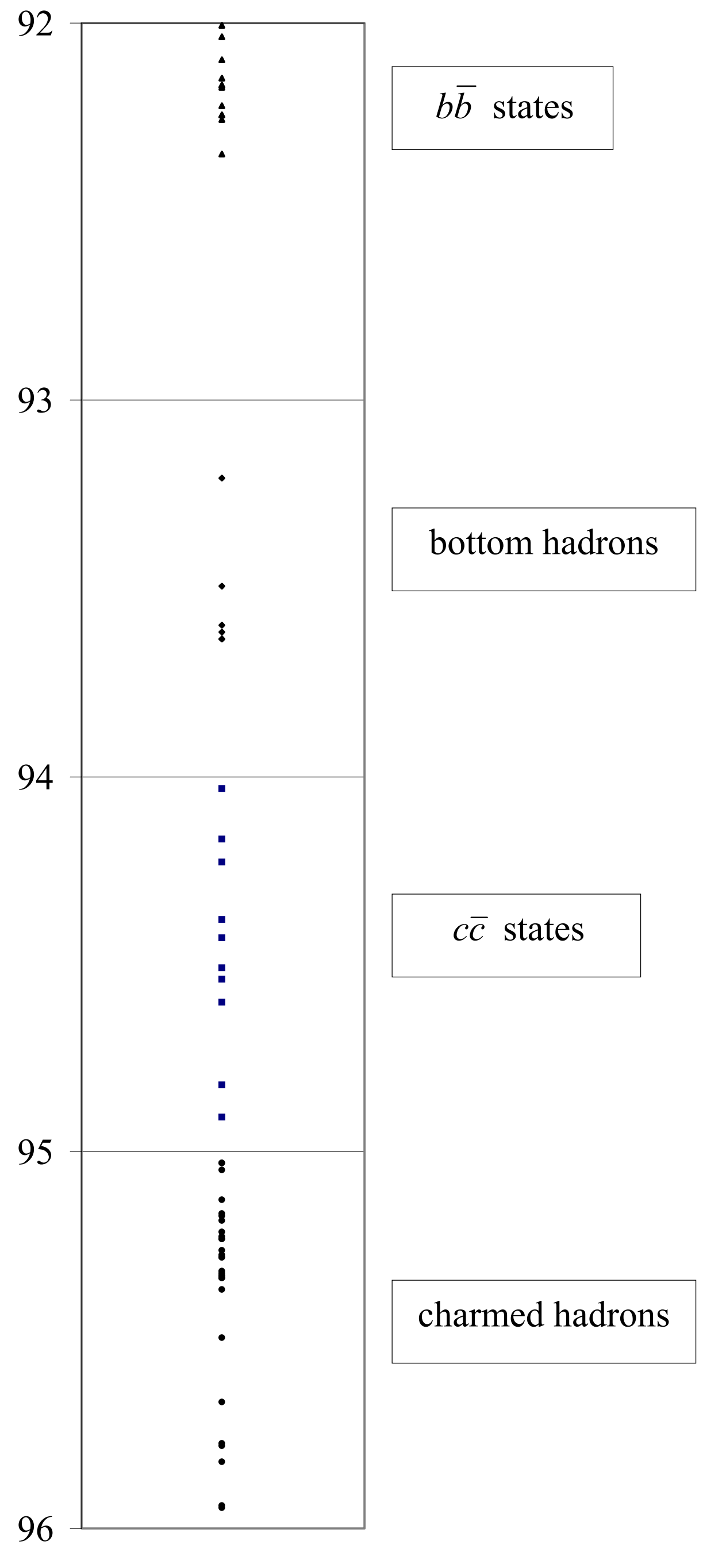

Figure 3: Mass level numbers, n, of all charmed and bottom hadrons, and $c\bar{c}$ and $b\bar{b}$ states from the Particle Data Group listings [13].

## 4 Baryon mass levels and mass construction

Values of n for the $J^P={}^1/_2{}^+$ and $J^P={}^3/_2{}^+$ baryon singlet states are shown in Figure 4. $\Sigma^0$ (uds, I=1, $I_3$=0) is included because it forms a 'mass-doublet' with $\Lambda$ (uds, I=0), centred on Mass Level 97. For the $m_{\Sigma^0} - m_\Lambda$ mass difference of 76.959 ± 0.023 MeV, n is equal to 103.0006 ± 0.0006. Other hadron mass differences will be shown to equal the masses of principal levels. The $\Lambda_c^+$ (udc) and $\Lambda_b^0$ (udb) baryons each appear to occupy first order mass sub-levels. $\Omega^-$ (sss) and $\Omega_c^0$ (ssc) apparently occupy fourth and third order sub-levels, respectively.

Mass level numbers, n, of the $J^P={}^1/_2{}^+$ baryons are presented in Figure 5, superimposed upon seventh order mass sub-levels. The isospin singlet states ($\Lambda$, $\Lambda_c^+$ and $\Omega_c^0$) occupy sub-levels of seventh order. The mass measurement uncertainty of $\Lambda_b^0$ (mass of 5624 ± 9 MeV) is too great to assign a sub-level of seventh order to that particle. The isospin doublets (p - n, $\Xi^0$ - $\Xi^-$ and $\Xi_c^0$ - $\Xi_c^-$) form mass-doublets centred on seventh order sub-levels. The masses of the isospin triplet states $\Sigma^+$, $\Sigma^0$ and $\Sigma^-$ are spread across two seventh order sub-level intervals and the masses of the isospin triplet states $\Sigma_c^{++}$, $\Sigma_c^+$ and $\Sigma_c^0$ cluster around a seventh order sub-level.

To a first approximation, the mass of baryon $q_1q_2q_3$ is produced by the addition of the masses of its valence quarks and of an electromagnetic quark interaction term to a neutral mass, $m_0$, arising from the strong interactions between the quarks [15]:

$$m_{q_1q_2q_3} = m_0 + m_{q_1} + m_{q_2} + m_{q_3} + EM\,term \qquad (5)$$

Inspection suggests that the masses of $\Lambda_c^+$ (udc) and $\Lambda_b^0$ (udb) are constructed along similar lines to (5) but from the mass of the neutral baryon, $\Lambda$ by the addition of the masses of the valence quarks:

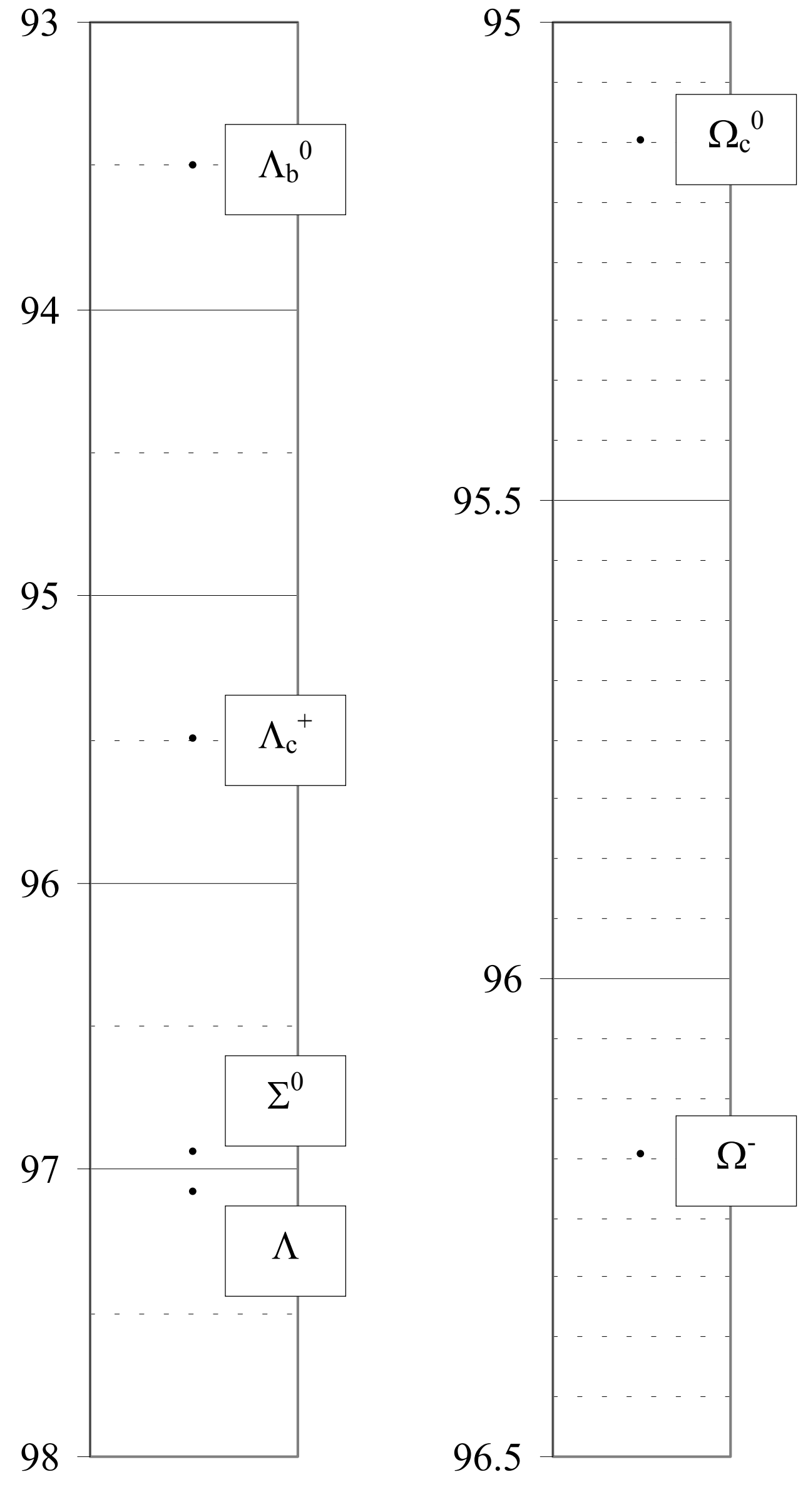

Figure 4: Mass level numbers, n, of the baryon singlet states. $\Sigma^0$ is included because it forms a mass-doublet with $\Lambda$.

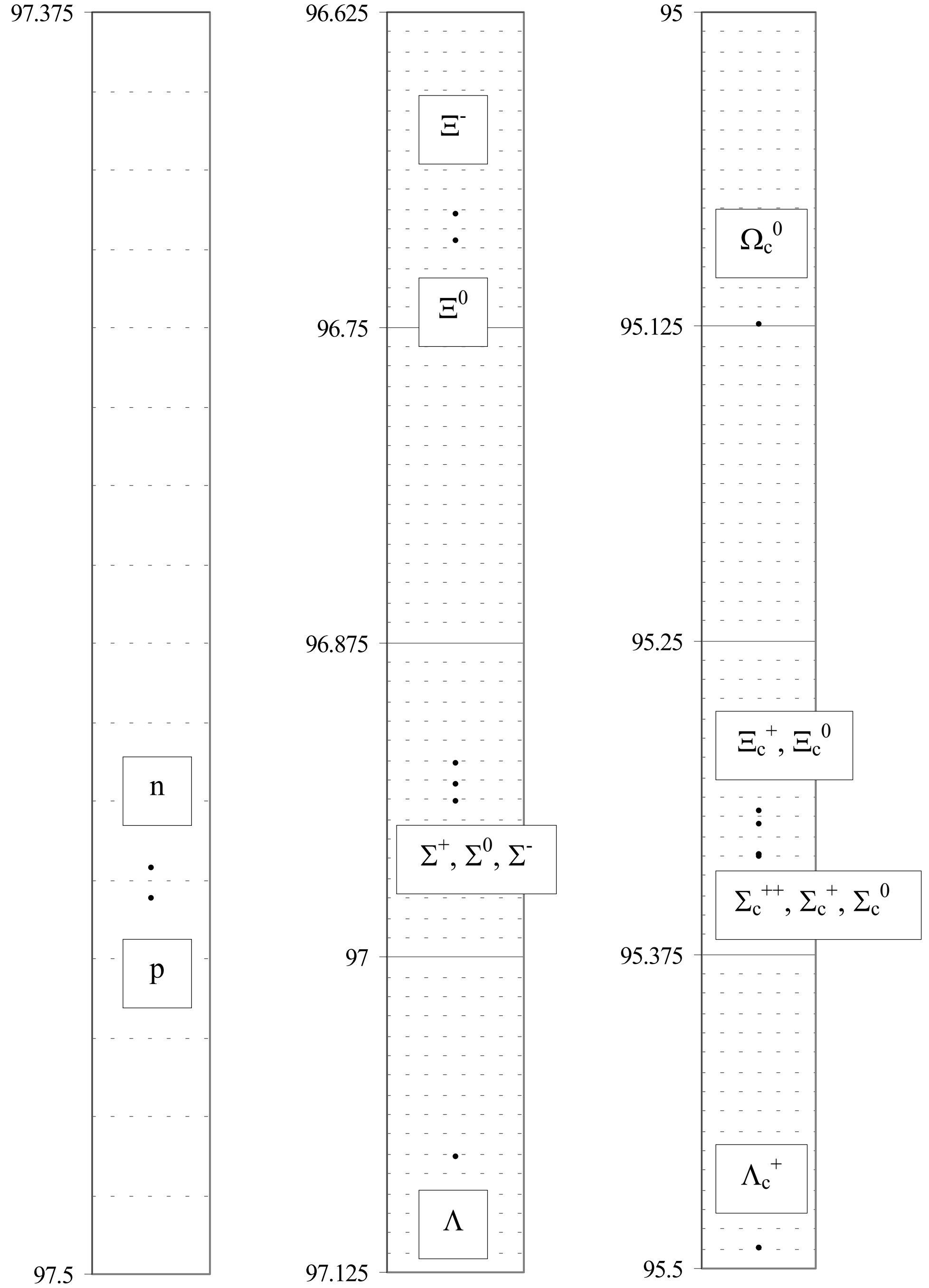

Figure 5: Mass level numbers, n, of the $J^P=½^+$ light and charmed baryons superimposed upon the seventh order sub-levels of the mass spectrum. Third order sub-levels are shown as solid lines.

$$m_{\Lambda_c^+} = m_\Lambda + m_u + m_d + m_c \qquad (6)$$

$$m_{\Lambda_b^0} = m_\Lambda + m_u + m_d + m_b \qquad (7)$$

The three baryons, $\Lambda$, $\Lambda_c^+$ and $\Lambda_b^0$, have identical quantum numbers $I(J^P)=0(^1/_2{}^+)$. (6) gives rise to a $\Lambda_c^+$ mass of 2283.34 MeV. The baryon apparently assumes the mass of the nearest seventh order sub-level to the mass value given by (6), which is 2284.54 MeV. The measured $\Lambda_c^+$ mass is 2284.9 ± 0.6 MeV. (7) gives rise to a $\Lambda_b^0$ mass of 5608.74 MeV. The nearest seventh order sub-level has a mass of 5617.04 MeV. The measured $\Lambda_b^0$ mass is 5624 ± 9 MeV.

A similar process of mass construction may apply to the $J^P={}^1/_2{}^+$ strange baryons. Here, in a manner which is repeated for $J^P={}^3/_2{}^+$ baryons and pseudoscalar mesons, mass construction produces an accurate value of mass for the neutral state of the isospin triplet and the most massive state of the isospin doublet. The masses of the neutral state, $\Sigma^0$ (uds), of the $\Sigma$ isospin triplet and the most massive state, $\Xi^-$ (dss), of the $\Xi$ isospin doublet appear to be constructed from the mass of a neutral precursor particle by the addition of the masses of their valence quarks, but the further addition or subtraction of the mass of the strange quark is required to balance the mass construction equations:

$$m_{\Sigma^0} = m_n + m_u + m_d + 2m_s \qquad (8)$$

$$m_{\Xi^-} = m_{\Sigma^0} + m_d + m_s \qquad (9)$$

(8) gives rise to a $\Sigma^0$ mass of 1192.71 MeV. The measured value is 1192.642 ± 0.024 MeV. (9) gives rise to a $\Xi^-$ mass of 1321.61 MeV. The measured value is 1321.31 ± 0.13 MeV.

The mass level numbers, n, of the $J^P={}^3/_2{}^+$ baryons are presented in Figure 6, superimposed upon seventh order mass sub-levels. The nearly mass-degenerate

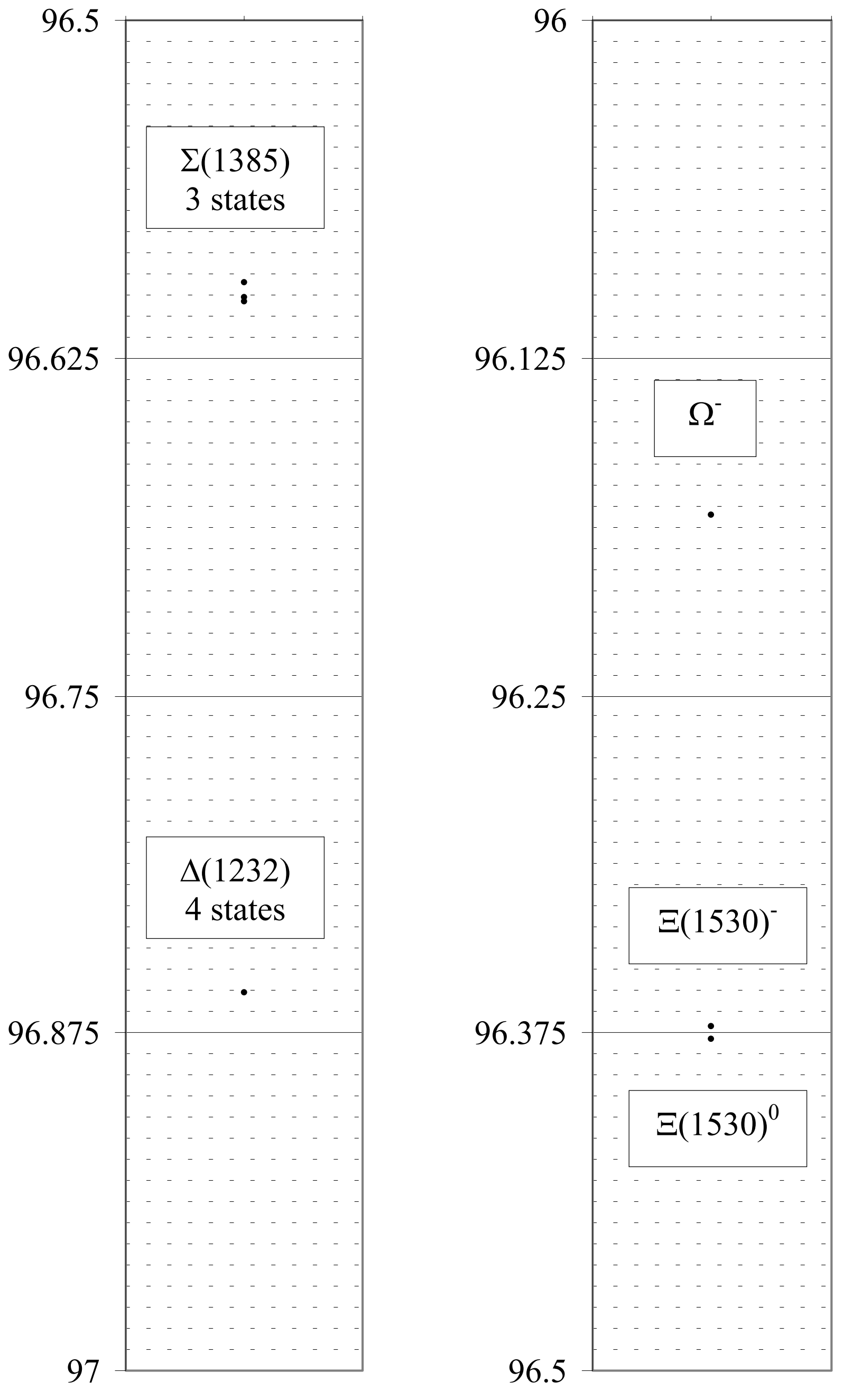

Figure 6: Mass level numbers, n, of the $J^P={}^3/_2^+$ light baryons superimposed upon the seventh order sub-levels of the mass spectrum. Third order sub-levels are shown as solid lines.

Δ(1232) states cluster around a sixth order sub-level, of mass 1232.16 MeV. The Σ(1385) isospin triplet states cluster around a seventh order sub-level and the Ξ(1530) isospin doublet states lie either side of a third order mass sub-level. The association of baryons with seventh and lower order mass sub-levels finally breaks down for the $\Omega^-$ baryon, which evidently occupies a higher order level.

The masses of the $J^P={}^3/_2{}^+$ light baryons increase by approximately equal increments as hypercharge increases in magnitude. Naturally, a relationship was sought that would describe this behaviour in terms of the unified model. In light of the findings for $J^P={}^1/_2{}^+$ strange baryons, the masses of the $J^P={}^3/_2{}^+$ strange baryons might be expected to develop from the masses of neutral precursor particles of identical spin. The mass of $\Sigma(1385)^0$ would be constructed from that of $\Delta(1232)^0$, the mass of $\Xi(1530)^-$ from that of $\Sigma(1385)^0$ and the mass of $\Omega^-$ from that of $\Xi(1530)^0$. With $m_{\Delta(1232)^0}$ = 1232.16 MeV (the mass of the associated sub-level), the following equations result:

$$m_{\Sigma(1385)^0} = m_{\Delta(1232)^0} + m_{101.500} \qquad (10)$$

$$m_{\Xi(1530)^-} = m_{\Sigma(1385)^0} + m_{101.501} \qquad (11)$$

$$m_{\Omega^-} = m_{\Xi(1530)^0} + m_{101.665} \qquad (12)$$

where $m_{101.500}$, $m_{101.501}$ *and* $m_{101.665}$ are the masses of sub-levels for which n = 101.500, n = 101.501 and n = 101.665, respectively.

The following mass construction equations are suggested for the $J^P={}^3/_2{}^+$ strange baryons:

$$m_{\Sigma(1385)^0} = m_{\Delta(1232)^0} + \left(\frac{\pi}{2}\right)^{\frac{1}{2}} m_s \qquad (13)$$

$$m_{\Xi(1530)^-} = m_{\Sigma(1385)^0} + \left(\frac{\pi}{2}\right)^{\frac{1}{2}} m_s \qquad (14)$$

$$m_{\Omega^-} = m_{\Xi(1530)^0} + \left(\frac{\pi}{2}\right)^{\frac{1}{3}} m_s \qquad (15)$$

which result in the following values of mass (measured values in brackets): $m_{\Sigma(1385)^0}$ = 1383.7 MeV (1383.7 ± 1.0 MeV), $m_{\Xi(1530)^-}$ = 1535.2 ± 1.0 MeV (1535.0 ± 0.6 MeV) and $m_{\Omega^-}$ = 1672.36 ± 0.32 MeV (1672.45 ± 0.29 MeV). The mass construction of a strange $J^P={}^3/_2{}^+$ baryon (Strangeness S) from a strange precursor particle is consistent with

$$m_{J^P=3/2^+\, strange\ baryon} = m_0 + \left(\frac{\pi}{2}\right)^{\frac{1}{|S|}} m_s \qquad (16)$$

## 5 Meson mass levels and mass construction

The $I(J^P)=0(0^-)$ and $I(J^P)=0(1^-)$ unflavoured states, η, ϕ, $\eta_c$, J/ψ and Y occupy mass sub-levels close to those of fifth order as shown in Figure 7. ρ (I=1) is included because it forms a mass-multiplet with ω (I=0), centred on third order sub-level 97.875. η′ occupies a high order mass sub-level.

The mass of η (547.30 ± 0.12 MeV) is coincident with that of a fifth order level, of mass 547.35 MeV. Like many other unflavoured meson states, ϕ, J/ψ and Y are closely associated with mass sub-levels of ninth order. Values of n corresponding to the masses of ϕ (1019.456 ± 0.020 MeV), J/ψ (3096.87 ± 0.04 MeV) and Y (9460.30 ± 0.26 MeV) are shown superimposed upon ninth order sub-levels in Figure 8. The $\eta_c$ mass (2979.7 ± 1.5 MeV) has too great a measurement uncertainty to assign a ninth order sub-level to that state. The masses of the states in Figure 8 have small uncertainties compared with the ninth order sub-level spacing, which, for each particle, is 0.0009 of the mass of the particle. Uncertainty in the value of the Planck Mass (1.2210 ± 0.0009 x$10^{19}$ GeV) gives rise to a systematic uncertainty in the positions of ninth order mass sub-levels equal to 0.38 of the level spacing.

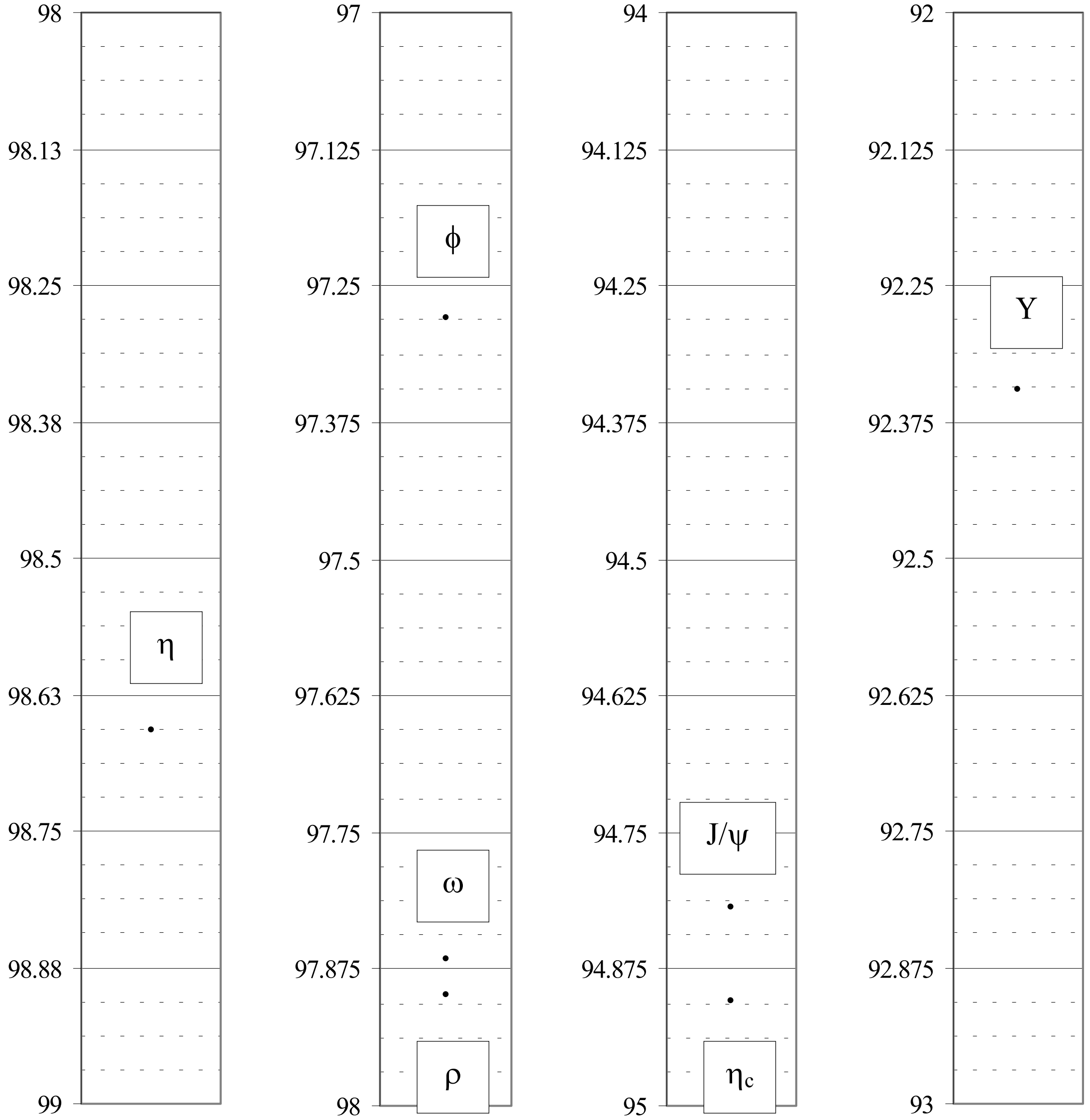

Figure 7: Mass level numbers, n, of the $J^P=0^-$ and $J^P=1^-$ unflavoured meson singlet states superimposed upon the fifth order sub-levels of the mass spectrum. Third order sub-levels are shown as solid lines. ρ is included because it forms a mass-multiplet with ω.

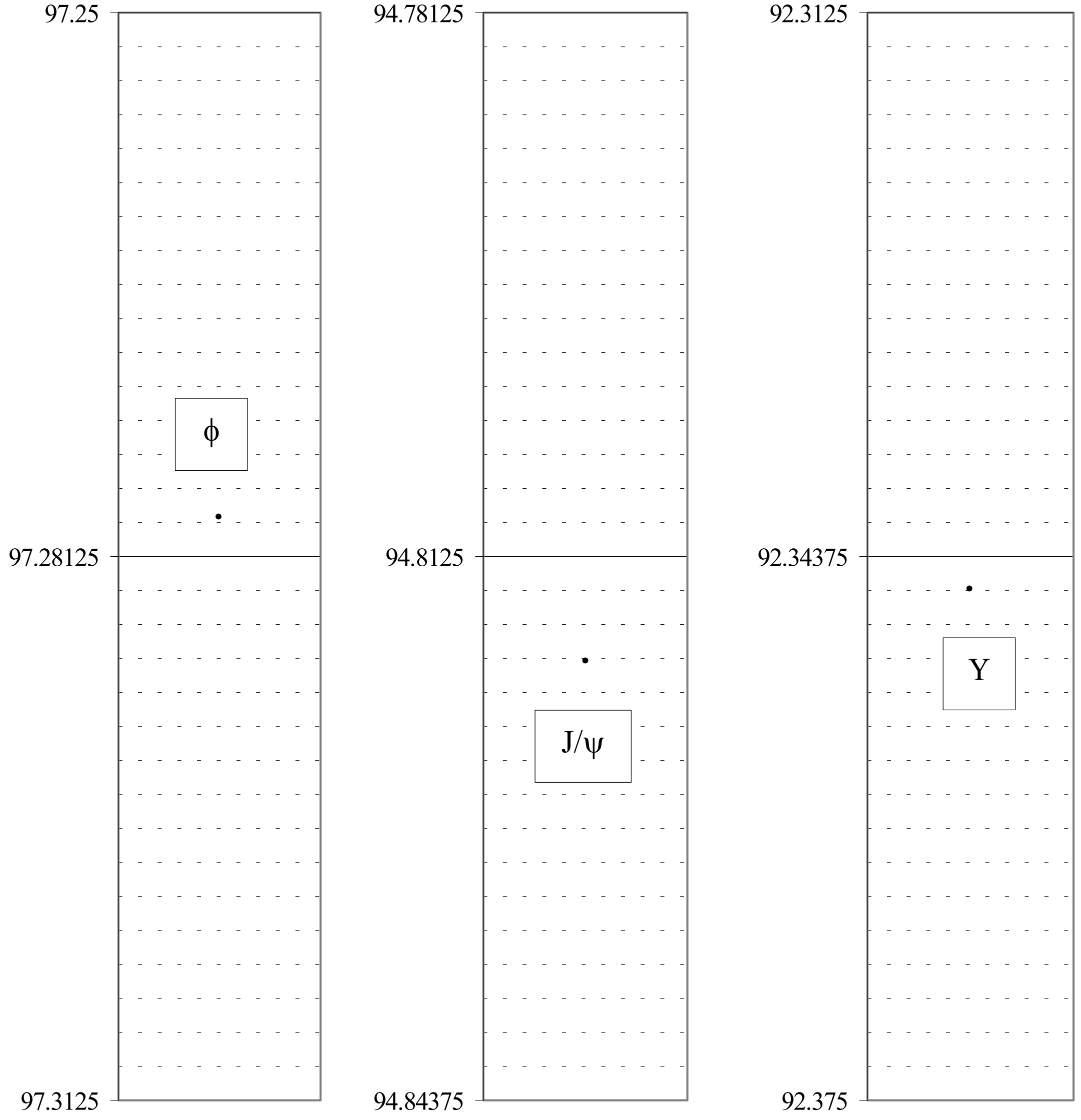

Figure 8: Mass level numbers, n, of unflavoured states ϕ, J/ψ and Y superimposed upon the ninth order sub-levels of the mass spectrum. Fifth order sub-levels are shown as solid lines.

The mass construction equations for the neutral vector states ϕ ($s\bar{s}$) and J/ψ ($c\bar{c}$) are:

$$m_{\phi} = m_{\omega} + 2m_s \qquad (10)$$

$$m_{J/\psi} = m_{\omega} + 2m_c \qquad (11)$$

The three states, ω, ϕ and J/ψ, have identical quantum numbers $I^G(J^{PC})=0^-(1^{--})$. (10) and (11) result in masses of 1024.40 MeV and 3095.26 MeV for ϕ and J/ψ, respectively. The measured masses are 1019.456 ± 0.020 MeV and 3096.87 ± 0.04 MeV, respectively. In both cases, the meson appears to have occupied a ninth order mass sub-level closer to the adjacent fifth order sub-level than the constructed mass. The nearest ninth order sub-levels to the measured masses have masses of 1019.32 MeV and 3097.07 MeV, respectively.

Values of n corresponding to the masses of all $b\bar{b}$ states from the Particle Data Group listings are shown superimposed upon ninth order mass sub-levels in Figure 9. All but the three heaviest states have small ($\leq 10^{-4}$) relative uncertainties in measured mass. Of the twelve states in the listings, eight are closely associated with ninth and lower order mass sub-levels. The remaining states occupy higher order sub-levels as suggested by the study of the χ states presented in Figure 10. The J=0 and J=2 states occupy sixth, seventh and eighth order mass sub-levels. The J=1 states apparently occupy higher order sub-levels.

Values of n for the pseudoscalar mesons and charged leptons are shown superimposed on fourth order mass sub-levels in Figure 11. $\pi^0$ and $\pi^{\pm}$ form a mass-doublet centred on a fifth order sub-level (not shown). $K^0$ and $K^{\pm}$ form a mass-doublet associated with a third order sub- level. $D^0$ and $D^{\pm}$ form a mass-doublet associated with a fourth order sub-level. The $B^{\pm}$ - $B^0$ isospin doublet and the $B_s^0$ meson apparently form a mass-multiplet associated with a level of third order. Each of the levels with which a mass-multiplet is associated is coincident with a fifth order sub-level. The meson singlet states of Figure 7 were also shown to be associated with sub-levels of this order. The charged leptons occupy high order mass sub-levels and will be shown to be related in mass to the pseudoscalar mesons.

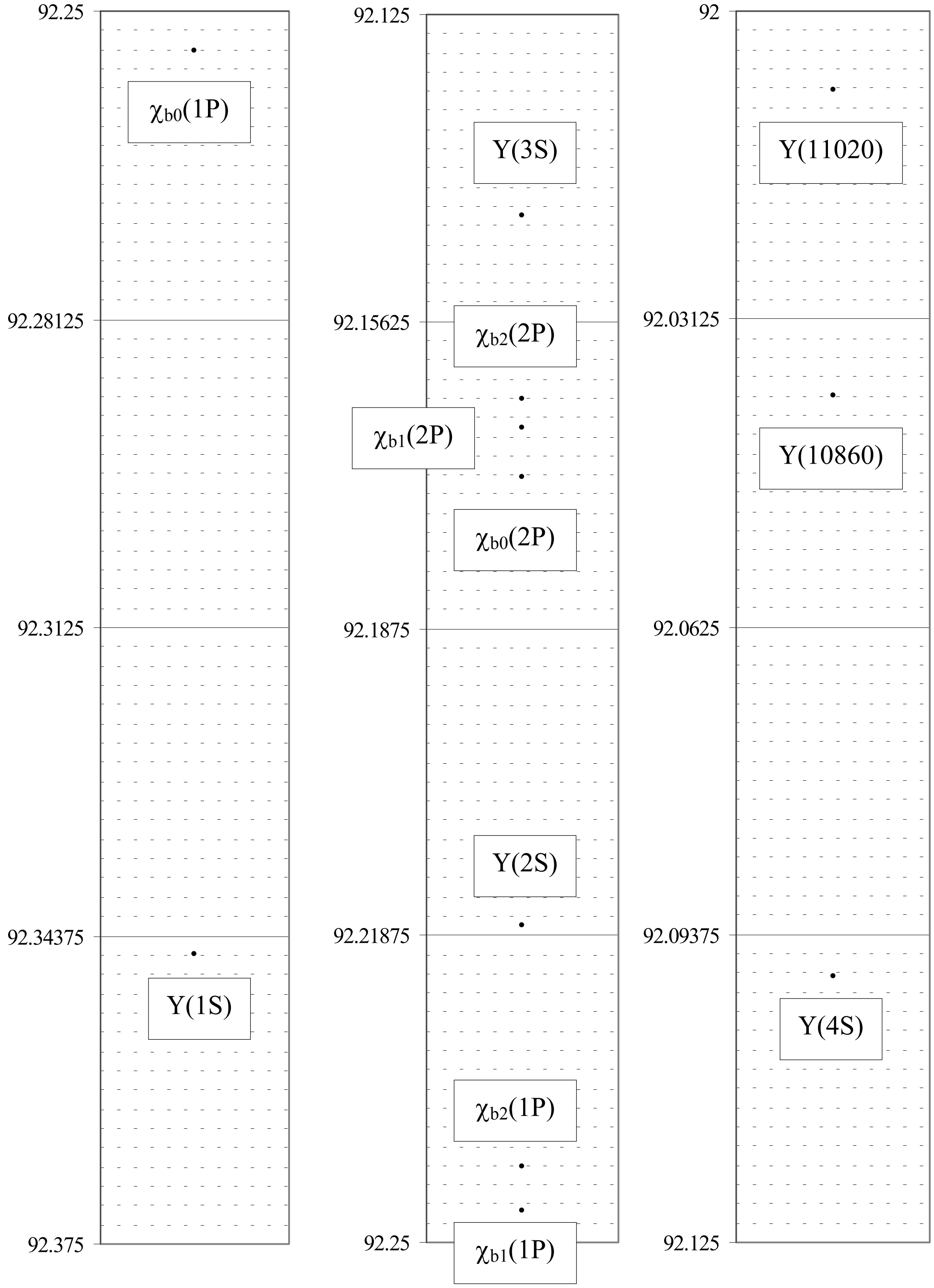

Figure 9: Mass level numbers, n, of the $b\bar{b}$ states superimposed upon the ninth order sub-levels of the mass spectrum. Fifth order sub-levels are shown as solid lines.

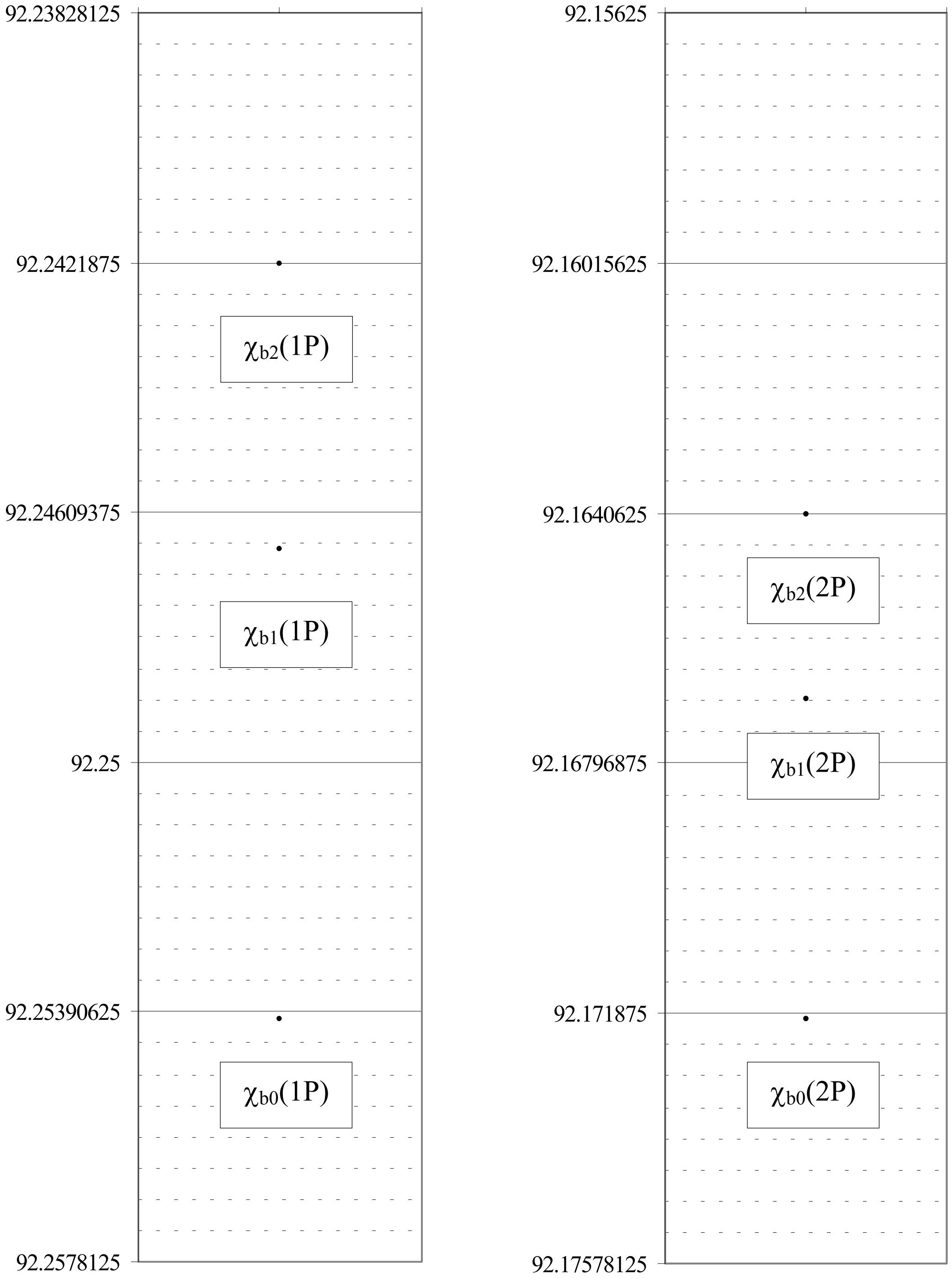

Figure 10: Mass level numbers, n, of the $b\bar{b}$ $\chi$ states. Eighth order sub-levels are shown as solid lines.

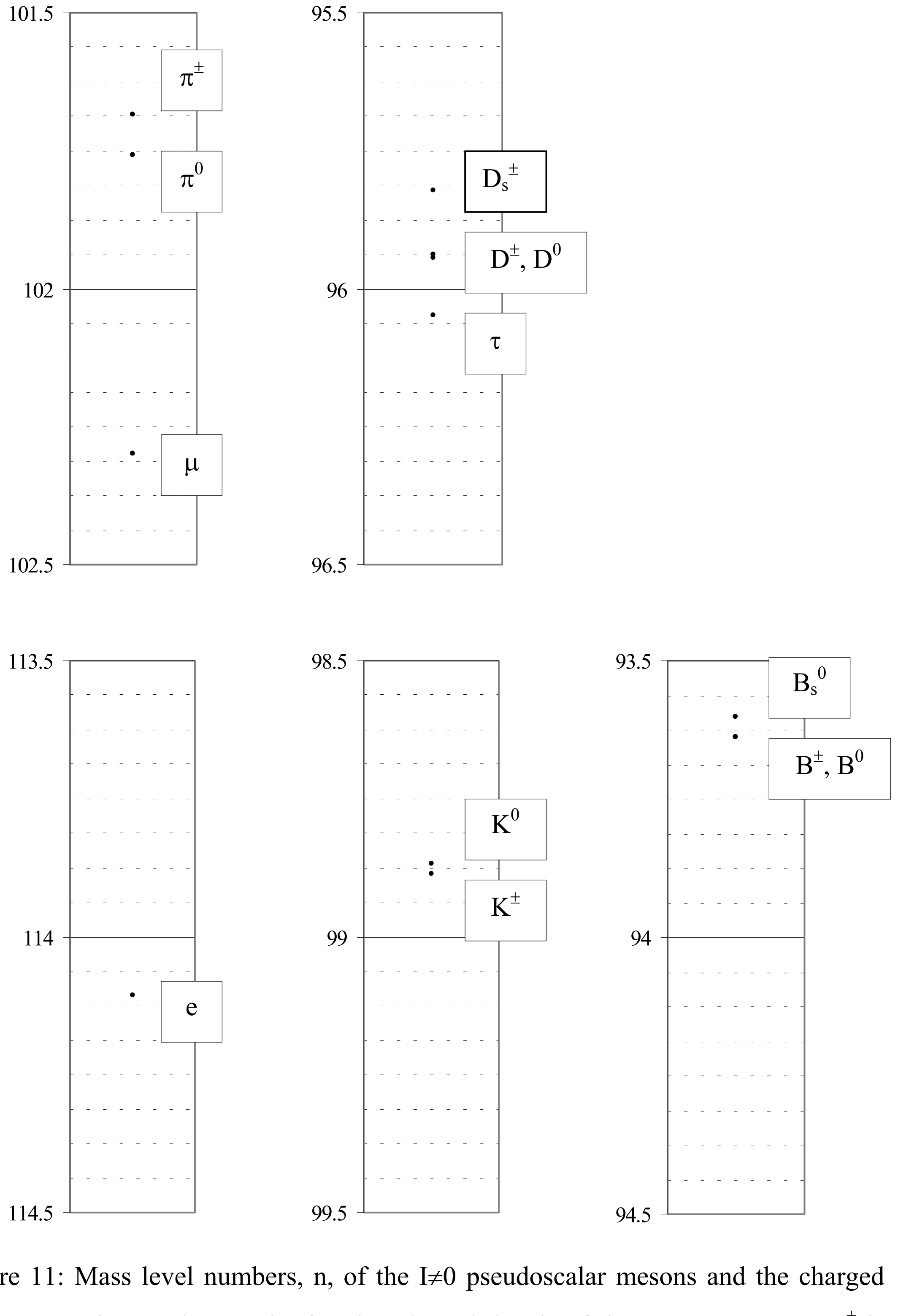

Figure 11: Mass level numbers, n, of the I≠0 pseudoscalar mesons and the charged leptons superimposed upon the fourth order sub-levels of the mass spectrum. $D_s^{\pm}$ is included because it appears to participate in a mass-multiplet with the D mesons and the tau lepton. $B_s^0$ also appears to participate in a mass-multiplet, with $B^{\pm}$ and $B^0$.

As in the mass construction of the $\Xi^-$ and $\Xi(1530)^-$ baryons, the masses of the heaviest particles, $K^0$ ($d\bar{s}$) and $D^\pm$ ($c\bar{d}$, $\bar{c}d$), in the K and D meson isospin doublets appear to be generated from the mass of a neutral precursor state. Here, the mass of each particle is consistent with the following mass construction equation:

$$m_{d\bar{q}} = m_{\pi^0} + \left|\frac{e}{Q_q}\right| m_q \qquad (12)$$

where e is the electronic charge and $Q_q$ is the charge of quark q. (12) gives rise to a $K^0$ mass of 497.73 MeV. The measured value is 497.672 ± 0.031 MeV. (12) also gives rise to a $D^\pm$ mass of 1869.49 MeV. The measured value is 1869.3 ± 0.5 MeV. The $K^\pm$ ($u\bar{s}$, $\bar{u}s$) and $D^0$ ($c\bar{u}$) mesons assume masses that largely reflect the $m_d - m_u$ quark mass difference of 4.79 MeV. The measured $m_{K^0} - m_{K^\pm}$ and $m_{D^\pm} - m_{D^0}$ mass differences are 3.995 ± 0.034 MeV and 4.78 ± 0.10 MeV, respectively.

The mass level numbers, n, of the I≠0 vector mesons are shown superimposed upon mass ladders in Figure 12. ω (I=0) is included because it apparently forms a mass-multiplet with ρ (I=1) associated with a third order sub-level. The $K^{*\pm}$ - $K^{*0}$ and $D^{*\pm}$ - $D^{*0}$ isospin doublets form mass-doublets associated with seventh and eighth order mass sub-levels, respectively.

## 4 Particle mass partnerships

Particle mass partnerships are defined as relationships between particles, or isospin multiplets, of different spin, isospin or isospin projection, in which the mass difference approximates closely to the mass of a principal level. Mass partnerships between hadrons normally involve particles with equal numbers of strange valence quarks, anti-strange quarks counting as strange quarks in this respect. In partnerships, the mass representative of an isospin doublet is that of its associated, interstitial, level. The mass representative of an isospin triplet is that of the neutral state. Values of n for several hadron, and multiplet, mass differences involving strange particles are presented in Table 3.

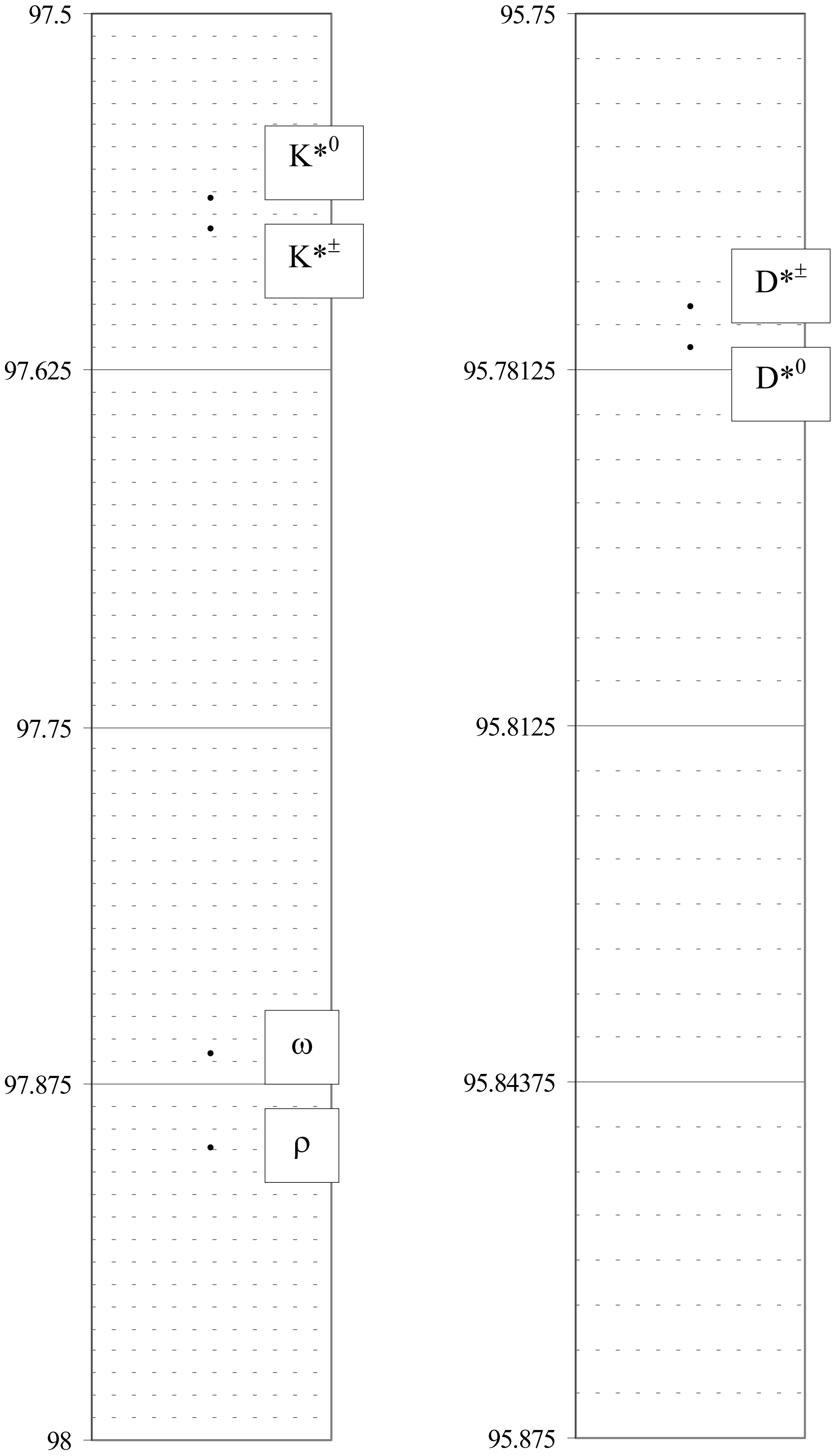

Figure 12: Mass level numbers, n, of the I≠0 light and charmed vector mesons. ω is included because it forms a mass-multiplet with ρ.

**Table 3 Values of n for hadron, and isospin multiplet, mass differences involving strange particles**

| Hadron or multiplet 1 | Hadron or multiplet 2 | Mass difference (MeV) | n |
|---|---|---|---|
| Λ (uds) | Σ (uus, uds, dds) | 76.959 ± 0.023 | 103.0006 ± 0.0006 |
| Σ (uus, uds, dds) | Σ(1385) (uus, uds, dds) | 191.1 ± 1.0 | 100.99 ± 0.01 |
| $\Sigma^+$ (uus) | $\Sigma^-$ (dds) | 8.08 ± 0.08 | 107.99 ± 0.02 |
| K* (us, ds) | Σ (uus, uds, dds) | 298.85 | 99.996 |
| φ ($s\bar{s}$) | Ξ (uss, dss) | 298.13 | 100.002 |

Baryon-baryon and vector meson-baryon partnerships occur.

The $\Xi^0$ - $\Xi^-$ isospin doublet has a mass difference of 6.48 ± 0.24 MeV, for which n is equal to 108.48 ± 0.08. The two baryons do not form a mass partnership, as defined here, though their mass difference is close to the mass of a first order sub-level. Some non-strange hadrons form partnerships in which the mass difference approximates to the mass of a principal level. The mass difference between the N(939) and Δ(1232) isospin multiplets is 293 MeV, for which n = 100.04. The $m_n - m_p$ isospin mass difference of 1.2933318 ± 0.0000005 MeV has an n-value of 112.05.

In Figure 11, the masses of the muon and the π mesons lie either side of Level 102 in a manner suggestive of a mass-doublet. The $m_\pi - m_\mu$ mass difference is 31.63 MeV, for which n = 104.97, suggesting that these particles may form a mass partnership.

The tau lepton and the D mesons lie either side of Level 96 in a manner suggestive of a mass-doublet. However, the mass-multiplet also appears to include the $D_s^{\pm}$ ($c\bar{s}$ , $\bar{c}s$ ) meson. The $m_{D_s^{\pm}} - m_{\tau}$ mass difference is 191.5 MeV, for which n = 100.98, suggesting that the tau lepton, the D mesons and the $D_s^{\pm}$ meson may form a mass partnership. Like the other pseudoscalar mesons and charged leptons, the K mesons and the electron lie close to mass levels (Levels 99 and 114, respectively) for which n is a multiple of 3. A mass partnership of sorts between the electron and the K mesons is suggested in Figure 11. The mass of the electron appears to have been displaced from a more symmetric position close to Level 99; with the electron in such a position, three meson-lepton mass partnerships would occur, associated with Levels 96, 99 and 102. The masses of the K meson isospin doublet and the electron are related through

$$m_K - \left(\frac{\pi}{2}\right)^{15} m_e = 48.96\ MeV\,(n = 104.002) \qquad (15)$$

suggesting that the electron and K mesons may form a type of mass partnership. Again, an almost exact principal level mass difference is found for a partnership involving strange hadrons. Curiously, the quantity $(\pi/2)^{15} m_e$ (of mass 446.91 MeV) is closely related to the mass of the K* vector meson isospin doublet seventh order sub-level (893.80 MeV):

$$\left(\frac{\pi}{2}\right)^{15} \frac{m_e}{m_{K*}} = 0.50001 \qquad (16)$$

The W and Z weak gauge bosons may also form a mass-doublet. The $m_Z - m_W$ mass difference of 10.764 ± 0.039 GeV has an n-value of 92.06, close to that of a principal level. The uncertainty in the measured mass of W (80.423 ± 0.039 GeV) is too great to assign a mass level to that particle. The mass of Z (91.1876 ± 0.0021 GeV), however, has a much smaller measurement uncertainty: this particle appears to occupy a sixth order sub-level, of mass 91.187 GeV.

## 5 Conclusions

All particles, including those with orbital angular momentum, occupy levels within a unified mass spectrum. The quarks occupy principal levels. The $J^P={}^1/_2{}^+$ and $J^P={}^3/_2{}^+$ light baryons, with the exception of $\Omega^-$, are associated with seventh and lower order mass sub-levels, as are the $J^P={}^1/_2{}^+$ charmed baryons. Baryon singlet states lie close to mass sub-levels of low order. Meson states with zero orbital angular momentum mostly occupy high order levels but are associated with levels of fifth and lower order.

The masses of many long-lived hadrons are constructed from the masses of neutral precursor particles of identical spin by the addition of quark masses. The mass construction of the K and D pseudoscalar mesons proceeds by the addition to the $\pi^0$ mass of a value of mass which is related to the mass-to-charge ratio of the flavoured valence quark. The masses of the strange $J^P={}^3/_2{}^+$ baryons are also constructed from the masses of neutral precursor particles of identical spin.

Mass partnerships exist between particles, and isospin multiplets, of different spin, isospin or isospin projection. Mass partnerships involving strange particles are often characterised by mass differences almost exactly equal to the masses of principal levels. Fermion-boson mass partnerships occur between $J^P={}^1/_2{}^+$ baryons and vector mesons, and between charged leptons and pseudoscalar mesons.

The mass of the electron appears to be displaced from those of the other particles and is related to that of the K* isospin doublet interstitial level.

The identification of a particle mass spectrum consistent with the experimental data on quark masses has opened the way to the discovery of hadron and lepton mass patterns and partnerships, and hadron mass construction equations. Hopefully, these findings will cast some light on the path to a unified theory of mass and force.